\documentclass{PoS}
\title{How reliable are bound-state parameters obtained from QCD sum rules?}
\ShortTitle{How reliable are bound-state parameters obtained from QCD sum rules?}
\author{Wolfgang Lucha
\\Institute for High Energy Physics, Austrian
Academy of Sciences, Nikolsdorfergasse 18, A-1050, Vienna,
Austria
\\E-mail: \email{Wolfgang.Lucha@oeaw.ac.at}}
\author{\speaker{Dmitri Melikhov}
\\Institute for High Energy
Physics, Austrian Academy of Sciences, Nikolsdorfergasse 18,
A-1050, Vienna, Austria, and
\\D.~V.~Skobeltsyn Institute of
Nuclear Physics, Moscow State University, 119991, Moscow,
Russia\\E-mail: 
\email{dmitri\_melikhov@gmx.de}}
\author{Silvano Simula\\INFN, Sezione di Roma III, Via della Vasca
Navale 84, I-00146, Roma, Italy\\ E-mail:
\email{simula@roma3.infn.it}}

\abstract{We discuss the extraction of ground-state parameters,
such as decay constants and form factors, from two- and
three-point dispersive sum rules, making use of a
quantum-mechanical potential model. This model provides a unique
possibility to probe the reliability and the accuracy of the
method of sum rules: one obtains the bound-state parameters by the
standard procedures adopted in sum rules, and compares these
results with the exact values, known in the potential model. We
demonstrate the presence of uncontrollable uncertainties in the
ground-state parameters obtained from sum rules and estimate their
magnitude. The uncontrolled uncertainties in ground-state form
factors are shown to be typically much larger than those in the
decay constants. In the example presented, the uncontrolled systematic error
in the extracted form factor is found to exceed the 10\% level.}

\FullConference{8th Conference Quark Confinement and the Hadron
Spectrum\\September 1-6, 2008\\Mainz. Germany}

\begin{document}
\section{Introduction}
A QCD sum-rule calculation of hadron parameters \cite{svz,ioffe}
involves two steps: (i) one calculates the operator product
expansion (OPE) for a relevant correlator and formulates the sum
rule which relates this OPE to the sum over hadronic states, and
(ii) one extracts ground-state parameters by a numerical
procedure. Each of these steps leads to uncertainties in the final
result.

The first step lies fully within QCD and allows for a rigorous
treatment of the uncertainties: the correlator in QCD is not known
precisely (because of uncertainties in quark masses, condensates,
$\alpha_s$, radiative corrections, etc.) but the corresponding
errors in the correlator may be controlled (at least in
principle).

The second step lies beyond QCD and is more cumbersome: even if
several terms of the OPE for the correlator were known precisely,
the hadronic parameters may be extracted from a sum rule only with
limited accuracy -- the corresponding error has to be treated as
a systematic error of the employed method.

In this talk, we present the results of our study of the
systematic errors of hadron parameters obtained from dispersive
sum rules. We address both the determination of the ground state's
decay constant from the two-point correlator \cite{lms_2ptsr} and
the extraction of the ground-state form factor from the
three-point correlator \cite{lms_3ptsr} in a quantum-mechanical
harmonic-oscillator (HO) potential model. This simple model has
strong advantages compared to more complicated cases: (i) The
bound-state parameters (masses, wave functions, form factors) are
known precisely. (ii) One can calculate the exact two- and
three-point functions. Thus, we may apply the standard
sum-rule machinery to extract the ground-state parameters and
compare them with the known exact values. In this way, we may
probe the reliability and the accuracy of the method.

We shall demonstrate that the standard procedures adopted in the
method of sum rules do not allow one to obtain rigorous error
estimates for the ground-state characteristics. Moreover, we show
that in the case of the form factors extracted from three-point
correlators, the uncontrolled systematic errors may be
considerably larger than those for the case of the decay constants
extracted from two-point correlators.

\section{Harmonic-oscillator model}
We consider a nonrelativistic HO model defined by the Hamiltonian
\begin{eqnarray}
H=H_0+V(r), \qquad H_0={\vec p}^{\,2}/2m, \qquad
V(r)={m\omega^2r^2}/{2}, \qquad r\equiv|\vec r\,|.
\end{eqnarray}
In this HO model, all characteristics of the bound states are
easily calculable. For instance, for the ground state ($n=0$) one
finds
\begin{eqnarray}
\label{E0} E_{0}=\frac{3}{2}\omega, \qquad R_{0}\equiv
|\Psi_{0}(\vec r=0)|^2=\left({m\omega}/{\pi}\right)^{3/2}, \qquad
F_{0}(q)=\exp(-q^2/4m\omega),
\end{eqnarray}
where the elastic form factor of the ground state is defined
according to
\begin{eqnarray}
F_{0}(q)=\langle\Psi_{0}|J(\vec q)|\Psi_{0}\rangle=\int d^3k\,
\psi^\dagger(\vec k)\psi(\vec k-\vec q), \qquad q\equiv |\vec q|,
\end{eqnarray}
with the current operator $J(\vec q)$ given by the kernel
\begin{eqnarray}
\label{J} \langle\vec r\,'|J(\vec q)|\vec r\rangle=\exp(i\vec
q\vec r)\delta^{(3)}(\vec r-\vec r\,').
\end{eqnarray}

\section{Polarization operator}
The basic quantity for the extraction of the ground-state wave
function $\Psi_{0}$ at the origin (i.e., of the decay constant) in
the method of dispersive sum rules is the correlator of two
currents \cite{svz}. Its quantum-mechanical analogue is
\cite{nsvz}
\begin{eqnarray}
\label{pi} \Pi(T)=\langle\vec r_{\rm f}=0|\exp(- H T)|\vec r_{\rm
i}=0\rangle.
\end{eqnarray}
For the HO potential, $\Pi(T)$ and its OPE (i.e., expansion at
small Euclidean time $T$) are known \cite{nsvz}:
\begin{eqnarray}
\label{piexact} \Pi(T) = \left(\frac{\omega
m}{\pi}\right)^{3/2}\frac{1}{\left[2\sinh(\omega T)\right]^{3/2}}
=\left(\frac{m}{2\pi T}\right)^{3/2}
\left(1-\frac{1}{4}\omega^2T^2+\frac{19}{480}{\omega^4 T^4}
+\cdots\right).
\end{eqnarray}

\section{Vertex function}
The basic quantity for the extraction of the form factor in the
method of dispersive sum rules~is the correlator of three currents
\cite{ioffe}. The analogue of this quantity in quantum mechanics
is \cite{lms_3ptsr}
\begin{eqnarray}
\Gamma(\tau_2,\tau_1,q) = \langle \vec r_{\rm f}=0|\exp(-H
\tau_2)J(\vec q)\exp(-H \tau_1)|\vec r_{\rm i}=0\rangle,
\end{eqnarray}
with the operator $J(\vec q)$ given by (\ref{J}). In the HO model,
the exact analytic expression for $\Gamma(\tau_2,\tau_1,q)$ was
obtained in \cite{lms_3ptsr}. The analogue of the OPE series, as
used in the method of sum rules in QCD, takes for equal times
$\tau_1=\tau_2=\frac{1}{2} T$ the following form:
\begin{eqnarray}
\label{gammaope}
&&\Gamma_{\rm OPE}(T,q)=\Gamma_0(T,q)+\Gamma_{\rm
power}(T,q), \qquad \Gamma_0(T,q)=\left(\frac{m}{2\pi
T}\right)^{3/2}\exp\left(-\frac{q^2T}{8m}\right), \nonumber\\
&&\Gamma_{\rm power}(T,q)=\left(\frac{m}{2\pi T}\right)^{3/2}
\left[-\frac{1}{4}\omega^2T^2+\frac{q^2\omega^2}{24m}T^3 +
\left(\frac{19}{480}\omega^4 - \frac{5q^4\omega^2}{1536
m^2}\right)T^4+\cdots\right].\qquad
\end{eqnarray}
In our actual computations, we take into account terms up to
O$(T^8)$ and O$(\omega^6)$ in $\Gamma_{\rm power}$. Notice that
the coefficients of each power of $T^n$ in the square brackets of
(\ref{gammaope}) are polynomials in $q^2$. This feature restricts
the applicability of three-point sum rules to the region of not
too large $q^2$.

\section{Ground-state parameters}
Making use of the standard assumption that the ground-state
contribution is dual to the low-energy region of the
free-quark diagrams, we obtain
\begin{eqnarray}
R_0 e^{-{E_0}T} &=& \int\limits_{0}^{z_{\rm
eff}(T)}dz\,\rho_0(z)\,e^{- z T}+\Pi_{\rm power}(T),
\\
R_0 F_0(q) e^{-{E_0}T} &=& \int\limits_{0}^{z_{\rm eff}(T,q)}dz_1
\int\limits_{0}^{z_{\rm eff}(T,q)}dz_2\,e^{-\frac12
z_1T}e^{-\frac12 z_2T} \Delta_0(z_1,z_2,q)+ \Gamma_{\rm
power}(T,q). \label{sr}
\end{eqnarray}
Here $\rho_0(z)$ and $\Delta_0(z_1,z_2,q)$ are the known spectral
densities of the two- and three-point Feynman diagrams of the
nonrelativistic field theory \cite{lms_2ptsr,lms_3ptsr}.

The relations (\ref{sr}) may be understood as the definition of
the exact $T$-dependent (and, in the case of the form factor,
also $q$-dependent) effective continuum thresholds, which are
obtained as the solutions of Eqs.~(\ref{sr}) for the exact
bound-state parameters on the l.h.s. The exact continuum
thresholds (specific for each quantity) may be calculated in the HO 
model where these bound-state parameters are known, but cannot be obtained directly from the OPE.
Obviously, the sum rules alone are {\em not\/} predictive: The ground-state 
parameters may be obtained in the method of
sum rules only if one imposes an independent criterion to fix the
effective continuum threshold.

\begin{figure}[hb]\begin{center}\begin{tabular}{cc}
\hspace{.14cm}\includegraphics[width=5.5cm]{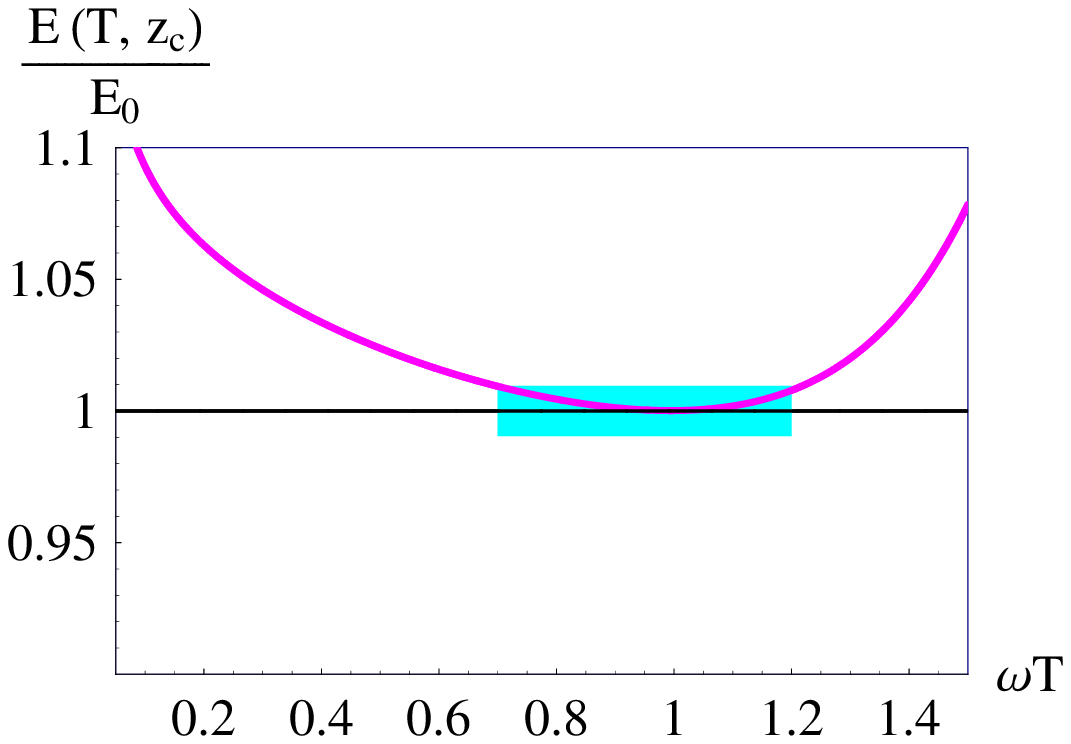}&
\includegraphics[width=5.5cm]{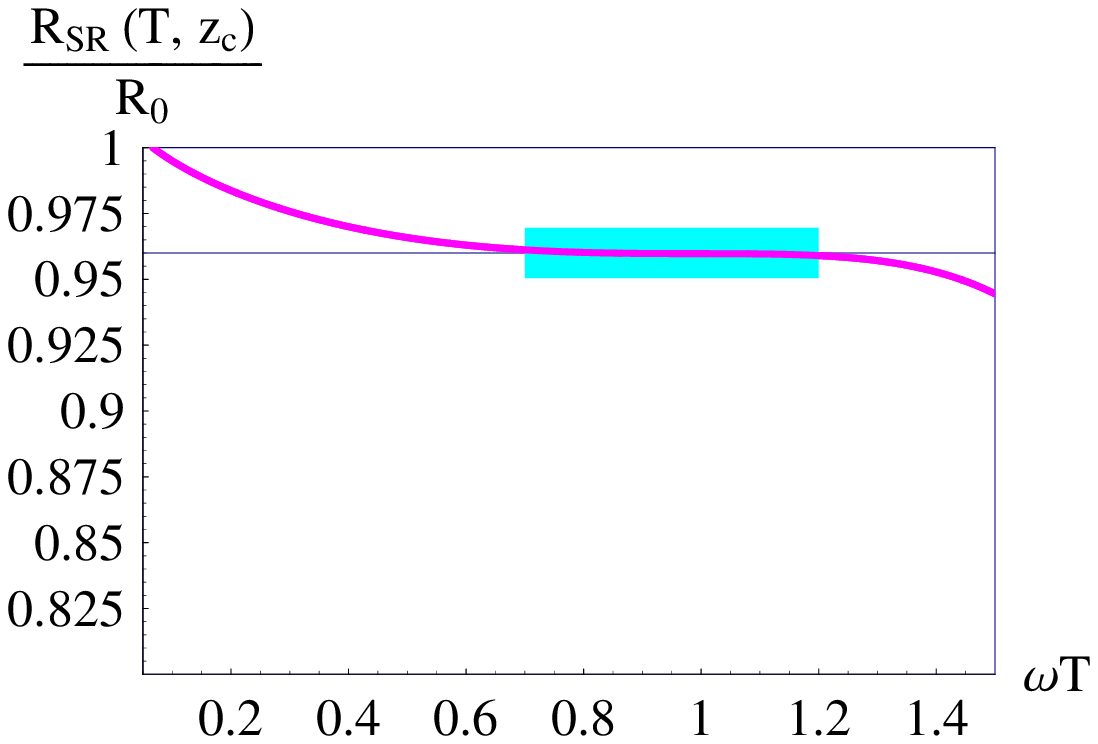}\\
(a)&(b)\\
\includegraphics[width=5.5cm]{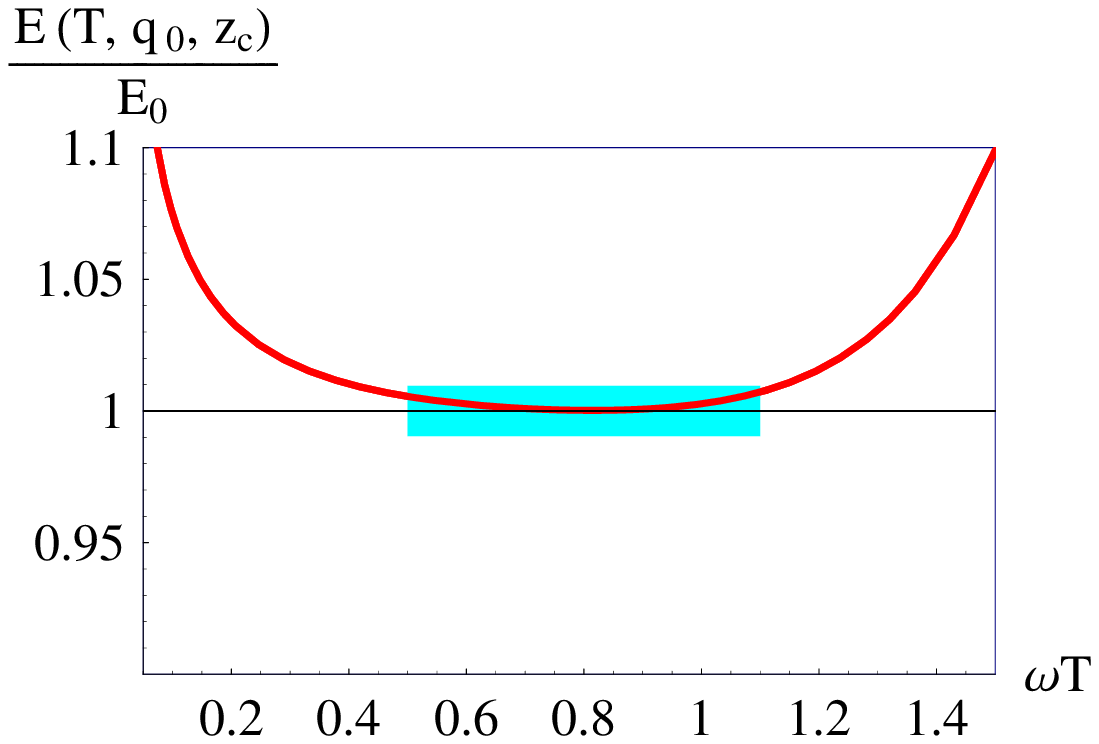}&
\includegraphics[width=5.5cm]{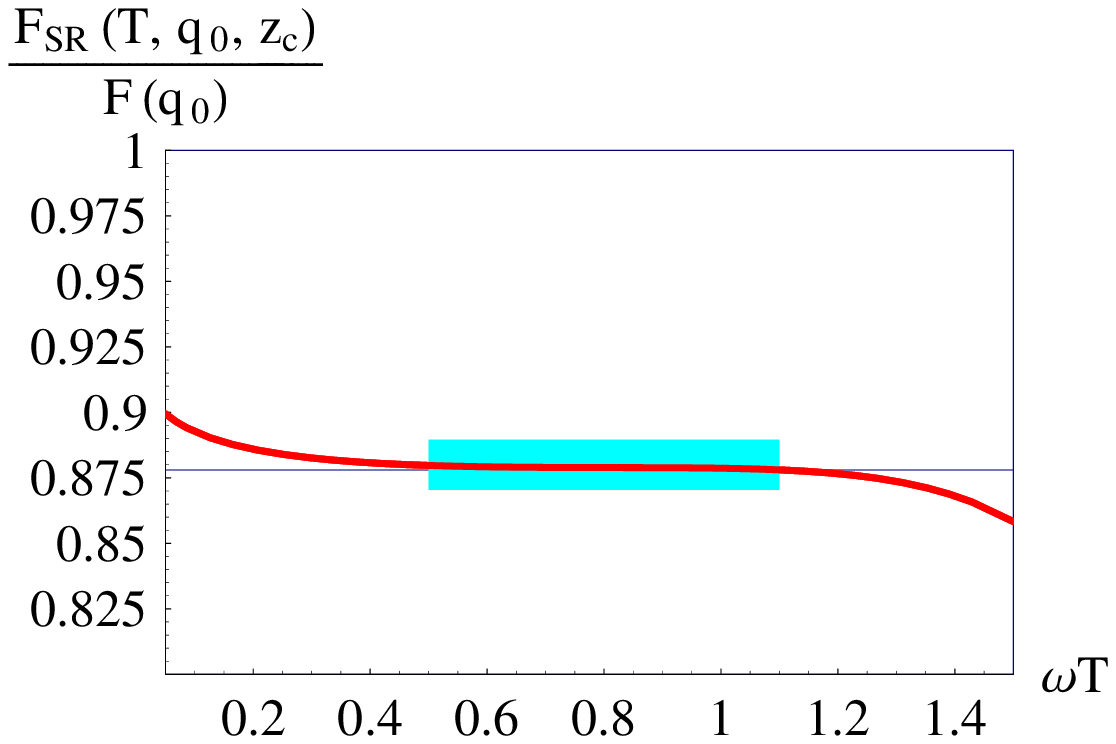}\\
(c)&(d)
\end{tabular}\end{center}
\caption{\label{plot:2.3} (a) The energy of the two-point cut
correlator $E(T,z_c)=-\partial_T\log \Pi(T,z_c)$ (where
$\partial_T$ is shorthand for $\partial_T\equiv\partial/\partial
T$) and (b) the sum-rule estimate for the parameter $R_0$ for
$z_c=2.454\,\omega$. (c) The energy of the cut correlator
$E(T,q_0,z_c)=-\partial_T\log \Gamma(T,q_0,z_c)$ and (d) the
sum-rule estimate for the form factor at $q_0=1.5\,\omega$ for
$z_c=2.42\,\omega$. The shaded rectangular areas indicate the
regions of stability with 1\% accuracy.}\end{figure}

Let us consider a restricted problem: namely, assume that the
energy $E_0$ of the ground~state is known and try to determine its
decay constant and elastic form factor from the sum rules
(\ref{sr}). The standard procedure adopted in this method is to
approximate the effective continuum threshold by a $T$-independent
quantity $z_c$ (for details, see \cite{lms_2ptsr,lms_3ptsr}).
Following this line, we obtained the results depicted in
Fig.~\ref{plot:2.3}. The parameter $R_{\rm SR}(T,z_c)$ extracted
from the sum rule is very flat in the stability region but
underestimates the known actual value by about 5\%. For the form
factor, the situation is even worse: for instance, at
$q_0=1.5\,\omega$, the form factor $F_{\rm SR}(T,q_0,z_c)$ is
perfectly flat in the stability region but, nevertheless, turns
out to be more than 10\% smaller than the known true value.

Let us emphasize the following dangerous point: (i) a perfect
description of the correlators with an accuracy better than 1\% in
the stability region, (ii) the deviation of the energy from
$E_0$~at~the level of only 1\%, and (iii) a very good stability of
the hadron parameters with better than 1\%~in~the stability region
lead, nevertheless, to an error of about 5\% for $R_0$ and more
than 10\% in the extracted value of $F(q_0)$. Clearly, these
errors could not be guessed on the basis of the other
numbers~found: the full picture mimics a very accurate extraction
of the hadron parameters, which is, however, \emph{not} true in reality!

\vspace{-.2cm}

\section{Summary and conclusions}
Let us summarize the lessons to be learnt from our analysis:

\noindent 1. The knowledge of the correlator in a limited range of
relatively small Euclidean times~(equivalently, large
Borel masses) is not sufficient for the extraction of the
ground-state parameters. In addition to the OPE for the
correlator, one needs an independent criterion for fixing the
effective continuum threshold.

\noindent 2. Assuming the effective continuum threshold to
be \emph{Borel-parameter-independent} allows one~to fix this
quantity by, e.g., requiring the average energy $E(T)$ to be close
to $E_0$ in the stability~region. In this case, however, the error
of the extracted ground-state parameter turns out to be typically
much larger than (i) the error of the description of the exact
correlator by the truncated OPE and~(ii) the variation of the
bound-state parameter in the Borel window.

\noindent 3. We stress that the stability of the
extracted ground-state parameter in the Borel window --- 
the standard criterion that is believed to control both the
reliability and the accuracy~of~the extracted ground-state
parameter --- does not, in fact, guarantee the extraction of its
true~value.

\noindent 4. The standard procedures for estimating the uncertainties 
of the extracted bound-state
parameters do not allow one to provide realistic error estimates.

The impossibility to control the systematic errors of
the extracted hadron parameters is the weak feature of the method
of sum rules and an obstacle for using the results from dispersive
sum rules in problems where rigorous error estimates are required.

Finally, we would like to comment on the obtained quantitative
estimates. In the HO model,~the ground state is well separated
from the first excitation contributing to the correlator by the
large~gap of $2\omega$. This feature renders the HO model a rather
favourable case for the application of sum rules. Whether or not a
comparable accuracy may be achieved in QCD --- where the
corresponding feature is absent --- is questionable.

\noindent {\it Acknowledgements.} The authors would like to thank
the organizers for arranging this very interesting meeting. D.~M.\
gratefully acknowledges financial support from the Austrian
Science Fund (FWF) under project P20573, and from the RFBR under
project 07-02-00551.

\end{document}